\documentclass [aps,prL,amssymb,amsmath,twocolumn,showpacs]{revtex4-1}

\usepackage{microtype}
\usepackage{amsmath}
\usepackage{amssymb}
\usepackage{graphicx}
\usepackage{verbatim}
\usepackage{bm}
\usepackage[dvipsnames]{xcolor}

\usepackage{graphicx,epsfig,amsfonts,amssymb}
\usepackage{hyperref}
  \hypersetup{colorlinks=true,citecolor=blue}

\usepackage{bm}
\usepackage{times}
\usepackage{lipsum}

\usepackage[all]{xy}

\newcommand{\nn}{\nonumber}

\newcommand{\balpha} {\bm \alpha}

\newcommand{\brho}{\bm{\rho}}

\newcommand{\bP} {\mathbf{P}}
\newcommand{\bp} {\mathbf{p}}

\newcommand{\ra}{\rangle}
\newcommand{\rar}{\rightarrow}
\newcommand{\da}{\downarrow}
\newcommand{\ua}{\uparrow}
\newcommand{\be}{\begin{eqnarray}}
\newcommand{\ee}{\end{eqnarray}}

\newcommand{\dg}{\dagger}

\newcommand{\hU} {\hat U}
\newcommand{\hM} {\hat M}
\newcommand{\hsigma} {\hat \sigma}
\newcommand{\hrho} {\hat \rho}
\newcommand{\bs}{\begin{equation}\begin{split}}
\newcommand{\es}{\end{split}\end{equation}}

\date{\today}

\begin{document}
\title{Phase Diagram of the Dynamics of a Precessing Qubit under Quantum Measurement}

\author{ Xinru Tang }
\author{ Fuxiang Li }
 \email{fuxiangli@hnu.edu.cn}

\affiliation{School of Physics and Electronics, Hunan University, Changsha 410082, China}

\begin{abstract}
We study the phase transitions induced by sequentially measuring a single qubit precessing under an external transverse magnetic field. Under projective quantum measurement, the probability distribution of the measurement outcomes can be mapped exactly to the thermodynamic probability distribution of a one-dimensional Ising model, whose coupling  can be varied by the magnetic field from ferromagnetic to anti-ferromagnetic. For the general case of sequential quantum measurement,we develop a fast and exact algorithm to calculate the probability distribution function of the ferromagnetic order and anti-ferromagnetic order, and a phase diagram is obtained in the parameter space spanned by the measurement strength and magnetic field strength.  The mapping to a long-range interacting Ising model is obtained in the limit of small measurement strength. Full counting statistical approach is applied to understand the phase diagram, and to make connections with the topological phase transition that is characterized by the braid group. This work deepens the understanding of phase transitions induced by quantum measurement, and may provide a new method to characterize and steer the quantum evolution. 

\end{abstract}
\maketitle

\microtypesetup{activate=true}

\section{Introduction}
\label{sec:intro}
Quantum measurement is one of the most intriguing properties in quantum mechanics \cite{QM01, QM02, QM03, QM04}. The understanding and utilization of quantum measurement is crucial to the future application of quantum information and quantum computation \cite{QC01},   quantum cryptography\cite{QC02}, and quantum sensing \cite{QC03}.  Quantum measurements can be either strong or weak, depending on the specific system and purpose of application . For  strong measurement, it is usually applied in initializing  and reading out quantum states  \cite{Elzerman2004Single,  Vamivakas2010Observation, Neumann2010Single, Morello2010Single, Jiang2009Repetitive, Bechtold2015, Bechtold2016}. For weak quantum measurement, it is particularly useful for monitoring quantum evolution \cite{Wiseman2010Q,  Goetsch94, Gurvitz1997, Korotkov1999, Liu2010} and maneuvering quantum state \cite{Korotkov2001, Korotkov2006, Blok2014}. More recently, great efforts have been devoted to the development of different experimental techniques   to detect the single qubit dynamics \cite{Jordan2005, Clerk2010, Maurer2012, Liu2017, Gefen2018, Balk2018, Cujia2019}, and to characterize the decoherence induced by environments \cite{Li2016, Wang2019, Sakuldee2019}. 

Continuous, or sequential quantum measurement has been theoretically studied using different approachs, like random walks in state space \cite{Oreshkov2005}, quantum Bayesian approach \cite{Jordan2006qubit}, and stochastic path-integral formalism \cite{Chantasri2013, Chantasri2015}.  Recently, it was theoretically discovered that the language of phase transition can be used to distinguish the 
weak and strong measurement, by mapping the probability distribution of sequential measurement outputs to the thermodynamic distribution of interacting Ising spin models \cite{Ma2018}. The authors find that, for a single qubit, or a two level system, under sequential quantum measurement, the boundary between weak and strong measurement can be very well defined by a critical value of measurement strength or duration. However, their study is mainly focused on a qubit without any dynamics. One would expect that there exist richer and more interesting phase transition behaviors if the qubit experiences its own dynamics besides that induced by measurement. Indeed, it was theoretically discovered that, in the presence of an external transverse magnetic field, the qubit dynamics may undergo a phase transition between coherent oscillation and quantum Zeno effect, induced by sequential weak measurement \cite{phase1996}. Furthermore, it was later found  that, this phase transition is associated with a topological transition that can be classified by different elements of the braid group \cite{Li2014EPL}.

In this paper, we study the interplay of external magnetic field and sequential quantum measurement on the dynamics of a single qubit, by mapping the measurement outcomes to the on-site  spin states of a one dimensional (1D) Ising spin model. We find that,  the presence of transverse magnetic field introduces an additional degree of freedom that induces the phase transition among the ferromagnetic, paramagnetic, and anti-ferromagnetic phases. We develop a fast and exact algorithm to calculate the  probability distribution of the ferromagnetic order and anti-ferromagnetic order,  and thus to determine the phase diagram in the parameter space spanned by field strength and measurement strength. In the limit of small measurement strength, the probability distribution can be mapped to a long range Ising spin model, which can help us understand the phase diagram. Moreover, using the full counting statistical approach, we can analytically obtain the probability distribution in the limiting cases of small measurement strength and field strength, which helps us make a connection with the topological phase transition discovered in Ref.~\cite{Li2014EPL}. Our findings provide deeper  understanding in the phase transitions induced by quantum measurements.

The paper is organized as follows. In Section II, we present the formalism that is needed to describe the dynamics of single qubit under quantum measurement and external transverse magnetic field. In Section III, we discuss the phase transition when the qubit is monitored by projective measurement. In Section IV, we develop the fast algorithm to calculate the probability distributions of ferromagnetic order and anti-ferromagnetic order, and obtain the phase diagram. We further find a long-range interacting Ising model than can capture physics in the case of small measurement strength. Furthermore, a full counting statistical approach is applied to obtain analytical expressions of probability distribution in the limiting case of small measurement strength and field strength. In Section V, the cases with different initial states and with nonzero relaxation rates are discussed. Conclusions are made in the last section.

\section{Formalism of measuring a precessing single quibit}
We consider the dynamics of a single qubit under a transverse magnetic field in the $\hat{y}$ direction. Its Hamiltonian is expressed as
\be
\hat{H}=\frac{1}{2}\hbar \omega_L \hsigma_y ,
\ee
in which $\hsigma_y$ is the $y$-component Pauli matrices, and $\omega_L$ the Larmor frequency. Without quantum measurement, the density matrix $\hat{\rho}$ that describes the quantum state of the quibit undergoes a unitary evolution, and it can be formally written as, 
\be
\hat{\rho}(t) = e^{-i{\hat H}t} \hat{\rho}_0 e^{i\hat{H} t},
\ee
with $\hat{\rho}_0$ being the intial density matrix, and $\hat{U}(t) = e^{-i\hat{H}t}$  the unitary evolution operator.  
In terms of Pauli matrices,  the single quibit density matrix can always be represented by four real parameters, $\rho_0$ and ${\bm \rho}\equiv \{\rho_{x}, \rho_ y,\rho_ z\}$:
\be
\hat{\rho} = \frac{1}{2}\Big[ \rho_0 \hat{1} + {\bm \rho} \cdot {\bm \hat{\sigma}}\Big].
\ee
 Starting from intial state vector $\{ \rho_0, \rho_x,  \rho_y,\rho_z \}$, the density matrix becomes, after revolution time $\tau$:
\be
\hat{\rho}(\tau) &=& \rho_0 \hat{1} + [\rho_x \cos(\omega) + \rho_z \sin(\omega)]  \hsigma_x \nn \\
&&+ [\rho_z \cos(\omega) - \rho_x \sin(\omega)]  \hsigma_z + \rho_y \hsigma_y. 
\ee
Here we introduce a parameter $\omega$ to describe the strength of external magnetic field:
\be
\omega= \omega_L \tau .
\ee
Together with the measurement strength, it will induce the phase transitions among ferromagnetic, paramagnetic and anti-ferromagnetic phases. 
Note that since the qubit precesses around the transverse magnetic field in the $y$-direction, the $y$-component of density matrix would not change, and thus can be set to be zero, $\rho_y(t) =0$. 

We adopt a sequential measurement scheme, described by a series of commuting POVM operators \cite{ Wiseman2010Q, Oreshkov2005} :
\be
\hat{M}_{\alpha} = \frac{1}{2}\Big (\sin \frac{\theta}{2} \hat{1} + \alpha \cos\frac{\theta}{2} \hsigma_z\Big) 
\ee
with $\alpha=\pm 1$ being the measurement outcomes. This measurement scheme allows us to consider both the weak and strong measurement with  adjustable measurement strength $\lambda = \sin\theta$ ranging from $0$ to $1$ when  $\theta$ ranges from $0$ to $\pi/2$. 
Under a single quantum measurement, the probability  of obtaining outcome $\alpha$ is given by
\be
P_{\alpha}=\hat{M}_{\alpha} \hat{\rho} \hat{M}_{\alpha}^{\dg},
\ee
and the normalized density operator after measurement becomes: 
\be
{\hat \rho}' =  \hat{M}_{\alpha} \hat{\rho} \hat{M}_{\alpha}^{\dg}/ P_{\alpha}.
\ee

After a series of such measurements with equal time interval $\tau$ and unitary evolution under transverse magnetic field $\omega_L$ in between, 
the combined evolution of density matrix can be formally expressed as: 
\be
{\hat \rho} = \hM_{\alpha_N} \hU_N\ldots  \hM_{\alpha_1}\hU_1 \hrho_0 \hU_1^{\dg} \hM_{\alpha_1}^{\dg} \ldots \hU_N^{\dg} \hM_{\alpha_N}^{\dg} /P_{\balpha}.
\ee
Here, $N$ is the total number of measurements, and $P_{\balpha}$ is the probability of obtaining a specific series of outcomes $\balpha = \{\alpha_1, \alpha_2, \ldots, \alpha_N \}$, given by:
\be
P_{\balpha} = {\rm Tr}[\hM_{\alpha_N} \hU_N\ldots  \hM_{\alpha_1}\hU_1 \hrho_0 \hU_1^{\dg} \hM_{\alpha_1}^{\dg} \ldots \hU_N^{\dg} \hM_{\alpha_N}^{\dg} ].
\ee

Since the $y$-component of density matrix can be set to be zero, we introduce a three component vector to describe the state of the qubit: ${\bf p} \equiv \{ \rho_0, \rho_z, \rho_x \}$. Starting immediately after the $(n-1)$-th measurement with vector ${\bf p}_{n-1}$, the state  experiences a precession $\omega$ of time $\tau$ and then is followed by the $n$-th measurement ${\hat M}_{\alpha_n}$. The new vector $\bp_n$  evolves in the following way: 
\be
{\bf p}_n = {\cal A}_{\alpha_n} {\bf p}_{n-1}
\ee
with the ``evolving matrix":
\be
{\cal A}_{\alpha} = \frac{1}{2}\left(
\begin{array}{ccc}
 1 & \alpha  \sin  \theta  \cos \omega  & -\alpha  \sin  \theta  \sin \omega  \\
 \alpha  \sin \theta  & \cos \omega & -\sin\omega  \\
 0 & \cos  \theta  \sin\omega  & \cos  \theta  \cos \omega  \\
\end{array}
\right) \label{eq:A} 
\ee
It contains $\alpha$ as the outcome, and two parameters $\theta$ and $\omega$ describing the measurement strength and the strength of transverse field, respectively. In this paper, we will utilize this evolving matrix to analyze the probability distribution of the outcomes and to determine the phase transitions. 
However,  before we discuss the general cases, we would like to, in next section, first  discuss the special case of projective quantum measurement, in which the probability distribution can be easily obtained and the mapping to Ising spin model is exact. 

\section{Phase transition induced by Larmor precession and projective measurement.}
For projective measurement with $\theta=\pi/2$, the measurement operator reduces to 
\be
\hM_{\alpha} = \frac{1}{2}[\hat{1}+ \alpha \hsigma_z]
\ee
Its effect on any initial wave function  is to collapse the wave function to become the eigenstate $|\alpha\ra$ of Pauli matrix $\hsigma_z$, depending on the outcome $\alpha=\pm 1$. In the language  of density matrix, the projective measurement operator $\hM_{\alpha}$ reduces the state $\hrho = (1/2)(\rho_0 \hat{1} + \brho\cdot {\bm \hsigma})$ to be $\hrho =  (1/2)( \hat{1} + \alpha \hsigma_z)$, with probability of obtaining outcome $\alpha$, $P_{\alpha} = (1/2)(\rho_0 + \alpha \rho_z)$.   Taking into account of the unitary evolution due to Larmor precession, one obtains the probability $P_{\alpha_0, \alpha}$ after one measurement with outcome $\alpha$ together with the previous outcome being $\alpha_0$: 
\be
P_{\alpha_0, \alpha} = \frac{1}{2} [1+ \cos(\omega)\alpha \alpha_0 ].
\ee


\begin{figure*}
 \scalebox{0.37}[0.37]{\includegraphics{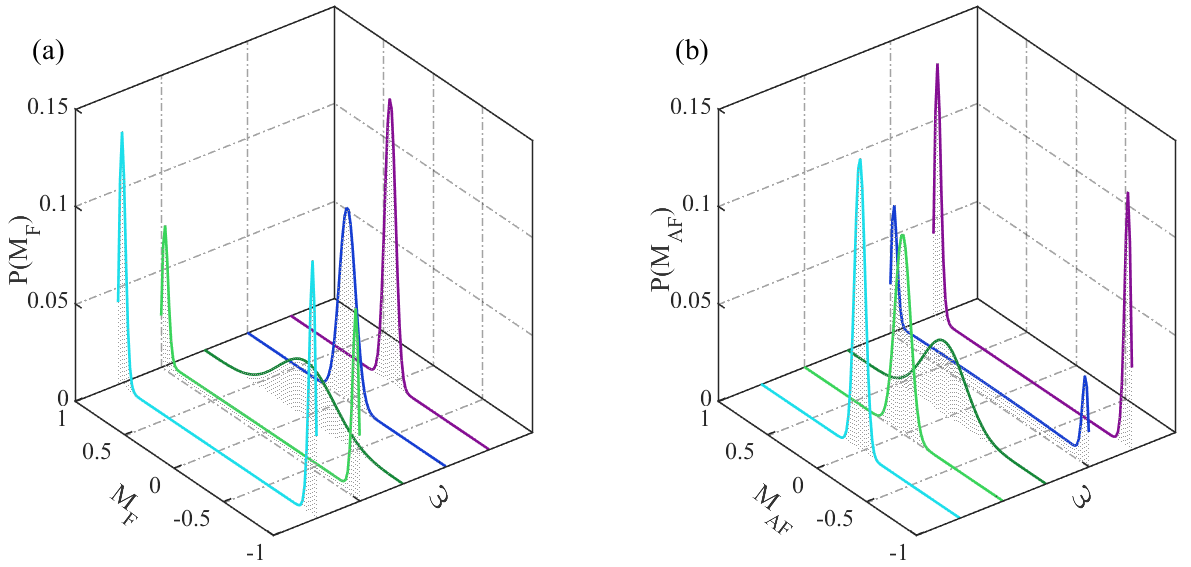}}
\caption{Phase transitions revealed by the probability distribution of ferromagnetic order $M_F$ (a) and anti-ferromagnetic order $M_{AF}$ (b). (a) shows the transition from the polarized  (PL)  phase with two peaks at $M_F\neq 0$, to the unpolarized (UPL) phase with only one peak centered  exactly at $M_F=0$, as one increases the value of $\omega$. From left to right, we fix $\theta = 2\pi /5$, and vary  $\omega$ to be $0.0$, $0.1$, $0.5$, $1.5$, $2.0$. (b) shows the transition from the  unpolarized phase with one peak centered at $M_{AF}=0$, to the anti-polarized (APL) phase with two peaks at $M_{AF} \neq 0$. Again, we fix $\theta = 2\pi/5$, and vary $\omega$ to be $1.1$,$1.6$, $2.5$, $3.0$, $\pi$ from left to right. } \label{fig:prob}
\end{figure*}

Therefore, the probability of obtaining a series of specific outcome $\balpha$ would be 
\be
P_{\balpha} = \frac{1}{2^N}\prod^N_{k=1}  [1+ \cos(\omega) \alpha_k \alpha_{k-1} ]
\ee
This probability can be mapped exactly to the thermodynamic probability of a nearest-neighbor coupled Ising spin model described by Hamiltonian:
\be
H =- J\sum_{k=1}^N  \alpha_k \alpha_{k-1}.
\ee 
The model is defined on a 1D lattice of  $N$ sites, with each site assigned with an Ising spin $\alpha_k =\pm 1$. 
The probability of finding a specific spin configuration is given by  the Gibbs distribution \cite{stat},  
\be
P_{\rm Ising} = \frac{1}{{\cal Z}}e^{-\beta H} = \frac{1}{{\cal Z}} e^{\beta J \alpha_k \alpha_{k-1}}.
\ee
Here, $\beta$ is the inverse of temperature $\beta=1/(k_B T)$. The partition function ${\cal Z} = {\rm Tr} e^{-\beta H} $ with the trace running over all the $2^N$ possible spin configurations.  
Setting $P_{\balpha}=P_{\rm Ising}$, we arrive at the following relation
\be
\tanh(\beta J) = \cos(\omega). 
\ee
This identity builds up the relation between the Larmor frequency of single qubit and the effective coupling of a 1-D Ising model. When $\omega = 0$, $\beta J=\infty$, corresponding a ferromagnetically coupled Ising spin chain  with infinite coupling strength $J$, or with finite coupling strength  $J$ but under zero temperature. In this case, the ferromagnetic phase is very well defined. As one increases $\omega$, the quantity $\beta J$ becomes finite, which can be understood as the increase of temperature to be nonzero, thus leading to the transition into a paramagnetic phase. As $\omega$ becomes $\pi$, $\beta J = -\infty$, corresponding to  anti-ferromagnetic coupling in the Ising model with infinite coupling strength, or finite strength but at zero temperature. In this case, the system is in the anti-ferromagnetic phase. 

Note that a similar discussion of phase transition among the ferromagnetic, paramagnetic and anti-ferromagnetic phases is made in Ref.~\cite{Ma2018} . However, this transition is induced by the angle between sequential measurements. This needs to change the measurement axis at each time of measurement, which requires experimental technique with  sufficiently high standards and precision. Otherwise, if, at each time of measurement, the angle with the previous measurement axis is not a constant, then it actually corresponds to introducing  disorder in the coupling strength of the Ising model.  From statistical mechanics, any amount of disorder would break the long-range ferromagnetic order in 1-D Ising model, making the phase transition difficult to be observed. In our case, however,  the phase transition is induced by the external magnetic field, which can be controlled in the experiment with high precision.

\section{Sequential measurement and phase diagram}

For the general cases of measurement strength $\theta$ and Larmor precession $\omega$, analytical approach becomes awkward and even impossible. In this section, we develop a fast algorithm which enables us to numerically and accurately determine the phase transitions. As is well known, to describe the magnetic phase transition, one needs to define a ferromagnetic order parameter $M_F$ and anti-ferromagnetic order parameter $M_{AF}$, and study their probability distribution, which would tell us about the information of phase transition, as was revealed in the Landau's theory of phase transition.   Using this algorithm, we determine the phase diagram in the parameter space spanned by $\theta$ and $\omega$. We show that this phase diagram can be quantitatively understood from the long-range Ising spin model and from the full counting statistical approach. 

\subsection{Recursion relation}
Specifically for the definition of  ferromagnetic order, we assume  $N$ measurements, or $N$ sites in the language of Ising model. Then we can define the ferromagnetic order as $M_F=(N_{\ua}- N_{\da})/N$, with $N_{\ua, \da}$ the number of sites with spin up (down), or the number of outcomes $\alpha=\pm 1$, respectively.   We define a probability $P(n_{\ua}, n)$ denoting the probability of obtaining $n_{\ua}$ outcomes of $\alpha=+1$ after $n$ measurements.  It is actually nothing but the first component of  state vector $\bP(n_{\ua}, n) \equiv \{\rho_0, \rho_z, \rho_x \}$, which describes the conditioned state vector after $n$-measurements and with $n_{\ua}$ outcomes of $\alpha=+1$.  Given the evolving matrix ${\cal A}_{\alpha}$ in Eq.~(\ref{eq:A}), the conditioned state vector $\bP(n_{\ua}, n)$ is given by the following recursion relation: 
\be
{\bf P}(n_{\ua},n+1) = {\cal A}_{+} {\bf P}(n_{\ua}-1, n) +  {\cal A}_{-} {\bf P}(n_{\ua}, n) \label{eq:PF}
\ee
The initial condition is simply $\bP(n_{\ua}, n=0) = \delta_{n_{\ua, 0}} \bp_0 $ with $\bp_0= \{ \rho_0(0), \rho_z(0), \rho_x(0)\}$ being the initial state vector. 
This recursion relation can be understood in the following way. Immediately after $(n+1)$-th measurement, the probability of obtaining $n_{\ua}$ number of up spins has two contributions: one is from the  previous probability ${\bf P}(n_{\ua}-1, n) $ of obtaining $n_{\ua}-1$ ups  together with the $(n+1)$-th measurement to be up (given by ${\cal A}_{+}$), the other is from the previous probability ${\bf P}(n_{\ua}, n) $ of obtaining $n_{\ua}$ ups together with the $(n+1)$-th outcome to be down (given by ${\cal A}_{-} $). 

After $N$ measurements, we come to a probability vector $\bP(N_{\ua}, N)$ describing the probability with $N_{\ua}$  outcomes of $\alpha = +1$. The first component  is just $P(n_{\ua}, n)$ that we are desired for, from which a symmetry breaking phenomena can be observed, as is illustrated  in Fig.~1(a), in which we plot this probability distribution as function of $M_F =(N_{\ua}- N_{\da})/N$ for different values of $\omega$ and $\theta$. Clearly a transition from two peaks to one peak is observed, as we fix the value of $\theta$ but  increase $\omega$.  The position of maximal probability transits from $N_{\ua }^{(max)} \neq N/2$ to exactly $N_{\ua}^{(max)}= N/2$, corresponding to the transition from nonzero ferromagnetic order $M_F\neq 0$ to exactly $M_F=0$.

For the definition of anti-ferromagnetic order parameter, we divide the $N$ sites into $N/2$ unit cells, with each unit cell consisting of two nearest neighbor sites. There are totally four cases of spin configurations $\{ \ua \ua\}$, $\{ \ua \da\}$, $\{ \da\ua\}$  and $\{ \da \da\}$ in one unit cell. Then we define the AFM order to be $M_{AF} = (N_{\ua\da} -N_{\da\ua})/(N/2)$, with $N_{\ua\da}$ ($N_{\da\ua}$) being the number of unit cells with the two neighboring spins in state $\{ \ua \da\}$ ( $\{ \da\ua\}$). 

Using similar procedure to the case of ferromagnetic order, we develop an algorithm to calculate the probability distribution of anti-ferromagnetic order $M_{AF}$. For this purpose,  we first define a quantity $n_A=n_{\ua\da} -n_{\da\ua}$,  with $n_{\ua\da}$ ($n_{\da\ua}$) being the number of unit cells with the two neighboring spins in state $\{ \ua \da\}$ ( $\{ \da\ua\}$).  Then we study the probability distribution $P_A(n_A, n)$ meaning the probability of obtaining $n_A$ after $2n$ measurements.  The recursion relation for the corresponding conditioned state vector $\bP(n_A, n)$ can be readily written as:
\be
\bP_A(n_{A}, n+1) =&&{\cal A}_{P} \bP_A(n_A, n)+   {\cal A}_{+}  {\cal A}_{-}\bP_A(n_A-1, n) \nn \\
 && + {\cal A}_{-}  {\cal A}_{+}\bP_A(n_A+1, n), 
\ee 
with the parallel measurement  operator given by
\be
{\cal A}_P = {\cal A}_+^2 + {\cal A}_-^2. 
\ee
It means that at $(n+1)$-th unit cell, or at $2(n+1)$-th measurements, the conditioned state vector $\bP_A(n_A, n+1)$ of obtaining $n_A$ is contributed from three sources, the first is from $\bP_A(n_A, n)$ with same number of $n_A$ together with the outcome of the $n$-th unit cell being in state $\{ \ua \ua\}$ or in state $\{ \da \da\}$, the second is from the probability $\bP_A(n_A-1, n)$ together with the outcome of $n$-th unit cell being in state $\{ \ua \da\}$  (contributing ${\cal A}_+ {\cal A}_-$), and the last is from  the probability $\bP_A(n_A+1, n)$ together with the outcome of $n$=th unit cell being in state $\{ \da \ua\}$ (contributing ${\cal A}_- {\cal A}_+$).

After $N$ measurements, we obtain the probability $P_A(N_A, N/2)$ which is just the probability distribution $P_A(M_{AF}, N)$ of the anti-ferromagnetic order  $M_{AF} =N_A/(N/2)$. We plot this probability distribution for different values of $\omega$ but with fixed $\theta$ in Fig.~1(b), and observe that there is indeed a transition from two peaks located at $M_{AF}\neq 0$ to one peak centered at $M_{AF} = 0$.

\begin{figure}[!htb]
 \scalebox{0.47}[0.47]{\includegraphics{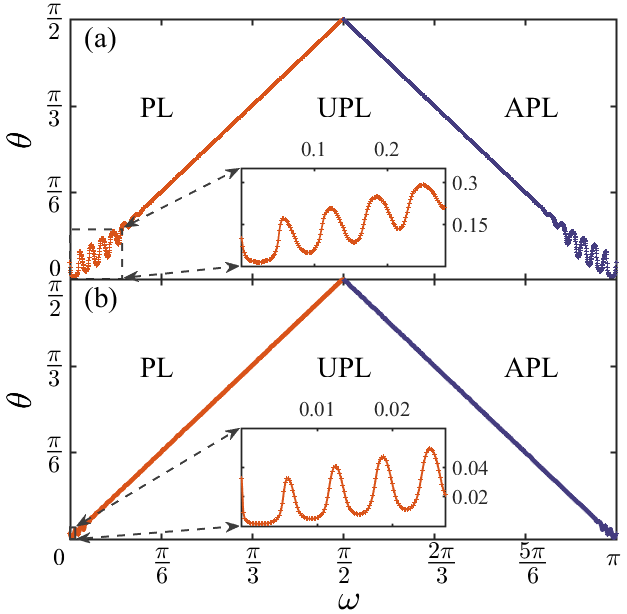}}
\caption{Phase diagram in the $\theta$-$\omega$ plane, for two cases of  $N=100$ (a) and $N=1000$ (b). The insets are the zoom-in of the oscillations that appear in the region of small $\theta$ and $\omega$.  Initial state is $\bp_0 =\{1, 0, 1\}$, and relaxation rate is zero.  }
\label{phase}
\end{figure}

\subsection{Phase diagram}

In order to quantitatively characterize these two transitions in terms of the probability distribution, we define three phases, polarized (PL) phase, unpolarized (UPL) phase, and anti-polarized (APL) phase, and obtain the phase diagram in the $\theta$-$\omega$ plane. From the probability distribution $P(M_F, N)$ of ferromagnetic order $M_F$, we can define the PL phase if the maximal probability is located at nonzero value  $M_F \neq 0$. From the probability distribution of $P_A(M_{AF}, N)$ of the anti-ferromagnetic order  $M_{AF}$, we can define the APL phase if the maximal probability is located at nonzero value of $M_{AF} \neq 0$. Otherwise, if both  the maximum of  $P(M_F, N)$ is located at $M_F=0$ and that of $P_A(M_{AF}, N)$ at $M_{AF} = 0$, then we call UPL phase. 

In Fig.~\ref{phase}, based on the calculations using the fast algorithm, we present the phase diagram for two different values of $N=100$ (a) and $N=1000$ (b). It is clear that, for a fixed measurement strength $\theta$, as one increases the Larmor precession $\omega$, the system undergoes in sequence the three phases, PL, UPL, and APL. There are two points worthy of noting. First, for small $N$, the finite size effect is obvious. Especially for the region of small $\theta$ and $\omega$, there appears an oscillation with a certain oscillation period. As shown in Fig.~2, the period of oscillation is about $0.063$ for $N=100$ (see the inset in Fig.~\ref{phase}a), and $0.0063$ for $N=1000$ ((see the inset in Fig.~\ref{phase}b). Actually, as will be discussed in next subsection, this oscillation behavior can be understood from a long-range interacting Ising model, and the period is found to be roughly $2\pi/(N-2)$, which agrees well with our numerical results.  Secondly, for large $N$, the finite size effect becomes diminished, and the phase boundary is almost a straight line, defined by $\omega=\theta$ for the PL/UPL phase boundary, and $\theta = \pi - \omega$ for the UPL/APL phase boundary. The boundary can be understood from analytical analysis by using the full counting statistical approach. 
 
\subsection{ Long-range interacting Ising model}
The phase diagram obtained by numerical calculation can be quantitatively understood by deriving a long-range interacting Ising model in the limit of weak measurement strength $\theta \ll 1$. From  Eq.~(\ref{eq:A}), one can obtain the final state vector $\bp_{N}$ after $N$ measurements with a specific series of outcomes $\balpha$ in a form like: 
\be
\bp_{N}(\balpha) = {\cal A}_{\alpha_N}\ldots {\cal A}_{\alpha_2}{\cal A}_{\alpha_1} \bp_0
\ee
In general, the mathematical form is too complicated to give rise to a compact analytical result. However, in the limit of $\theta \ll 1$, one is lucky to find that the probability is given by 
\be
P(\balpha) \sim e^{ \theta^2  \sum_{j<k}^N \cos[(k-j -1)\omega] \alpha_j \alpha_k}
\ee
This probability distribution can be recognized as the Gibbs distribution of  a long-range interacting Ising model with Hamiltonian: 
\be
H = - \theta^2 \sum_{j<k}^N \cos[(k-j -1)\omega] \alpha_j \alpha_k
\ee

For the case of $\omega =0$, this model reduces to the long-range ferromagnetic Ising Hamiltonian  obtained in Ref.\cite{Ma2018}. More interesting cases occur with increasing $\omega$ when the long range couplings gradually changes from ferromagnetic  to anti-ferromagnetic. The longest-range coupling is between the site $1$ and $N$ with coupling strength $J_{1,N} =\theta^2 \cos[(N-2)\omega]$. As $\omega$ increases, this coupling strength $J_{1, N}$ starts to oscillates, first decreases from positive to negative and then increases back to positive. The oscillation period is given by  $\delta_{\omega} = 2\pi / (N-2)$.  The other shorter-range couplings also oscillate with $\omega$, but with a smaller period. Totally, this picture gives rise to an oscillating behavior on the phase boundary. However, at larger $\omega$, the coupling strengths of different range oscillating  with different periods interfere with each other and the amplitude of  oscillation in the phase boundary finally diminish, as is revealed in the phase diagram in Fig.~\ref{phase}.

\subsection{ Full counting statistical approach}
In this subsection, we would like to understand the phase diagram by using the full counting statistical approach, in order to obtain  analytical expressions for the probability distribution functions. Define the generating function for the conditioned state vector $\bP(n_{\ua}, n)$ at the $n$-th measurement \cite{Noise},
\be
{\bf Z}(\chi, n) = \sum_{n_{\ua}=0}^{\infty} {\bf P}(n_{\ua}, n) e^{i \chi n_{\ua}}
\ee
The recursion relation (\ref{eq:PF}) becomes: 
\be
{\bf Z}(\chi, n) = ({\cal A}_+ e^{i\chi} + {\cal A}_-) {\bf Z}(\chi, n-1) 
\ee
Through this method, the generating function ${\bf Z}(\chi, N)$ after $N$ measurements would be readily written down. Thus the probability $P(N_{\ua} , N)$ of obtaining $N_{\ua}$ after $N$ measurements would be analytically calculated from the generating function: 
\be
P(N_{\ua}, N) = \frac{1}{2\pi}\int_0^{2\pi} d\chi e^{-i \chi N_{\ua}} Z_0(\chi, N) \label{eq:P0}
\ee

In general cases,  the analytical expression for $Z_0(\chi, N)$ is difficult to obtain. However, for the limiting case with small $\omega$ and small $\theta$, one can obtain a closed form. 
Indeed, in this limit, the ``Hamiltonian" ${\hat K} ={\cal A}_+ e^{i\chi} + {\cal A}_- $ governing the dynamics of ${\bf Z}(\chi, n)$ reduces  to 
\be
{\hat K}(z) = \frac{1}{2} \left(
\begin{array}{ccc}
 z+1 &   (z-1)\theta & 0 \\
   (z-1)\theta & z+1 & -(z+1)\omega \\
 0 &  (z+1) \omega & z+1 \\
\end{array}
\right)
\ee
with $z=e^{i\chi}$. 
This ``Hamiltonian" has three eigenvalues: 
\be
E_{1,2} = \frac{1}{2}\Big[(z+1) \pm \varepsilon(z) \Big] , ~~ E_3= \frac{1}{2} ( z+1),
\ee
with 
\be
\varepsilon(z) =  \sqrt{(z-1)^2\theta^2 - (z+1)^2 \omega^2}
\ee

It is interesting to note that, for the first two eigenvalues, if we set $z=0$, they reduce to $E_{1,2}=\frac{1}{2}  (1 \pm \sqrt{\theta^2 - \omega^2})$, which are  real for $\theta> \omega$, but become imaginary for $\theta< \omega$, leading to quite distinguished behaviors of ${\bf Z}(\chi, t)$ in the large $N$ limit, and thus that of probability distribution function $P(N_{\ua}, N)$. Generally, the generating function $Z_0(\chi, N)$ can be written as $Z_0(\chi, N) = \sum_{j=1,2,3} c_j (\chi) E_j^N$, with $c_{j}$ being the coefficients that are dependent on the initial states. From a similar equation, we remind that in Ref.~\cite{Li2014EPL}, an interesting topological phase transition was identified, described by a braiding group in the space of complex eigenvalues as functions of $\chi$ ranging from $0$ and $2\pi$. Here, we want to argue that, the ferromagnetic-paramagnetic phase transition may provide another point of view for this transition, whose transition line is also defined by $\omega = \theta$.  

Indeed, for simplicity, we consider the case with initial state $\hrho_0 = \frac{1}{2} (\hat{1} + \hsigma_x) $, corresponding to initial condition for ${\bf Z}(\chi, n=0) = (1, 0, 1)$.  The evolution can be obtained explicitly: 
\be
 Z_0(z, N) = f_z  E_3^N + \frac{1}{2} (1-f_z) \Big( E_1^N + E_2^N\Big) \label{eq:Z0}, 
\ee
with 
\be 
f_z =  \frac{ (z+1) \omega} { (z-1)\theta  +(z+1) \omega }.
\ee
In obtaining the probability distribution from Eq.~(\ref{eq:P0}), one can make a variable change: $z=e^{i\chi}$, thus transforming the integration to be a contour integration on the unit circle ${\cal C}$ in the complex plane: 
\be
P(N_{\ua}, N) = \frac{1}{2\pi i}\oint_{\cal C }dz \frac{1}{z^{N_{\ua}+1}} Z_0(z, N)
\ee
In the large $N$ limit, we can use the stationary phase approximation to obtain analytical results. 
We discuss the two limiting cases with $\theta \gg \omega$ and $\theta \ll \omega$ . 

\begin{figure}[!htb]
 \scalebox{0.55}[0.55]{\includegraphics{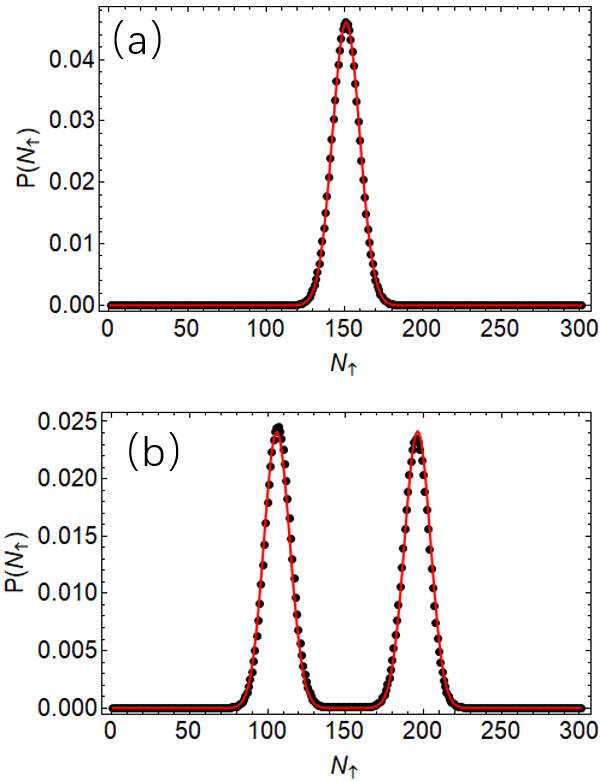}}
\caption{The probability distribution obtained from the analytical results (red lines) are compared with the exact numerical calculations (black dots). Distribution function (\ref{eq:p1})  is used  in (a) with parameters: $\theta=0.01$ and $\omega=0.2$, while Eq.~(\ref{eq:p2}) is used in (b) with $\theta=0.3$ and $\omega=0.001$. }
\label{distribution}
\end{figure}

For the case with $\theta \ll \omega$,  $f_z \rar 1$,  thus the second term in Eq.~(\ref{eq:Z0}) vanishes compared to the first term. Then we have
\be
P(N_{\ua}, N) =  \frac{1}{2\pi i}\oint_{\cal C }dz \frac{1}{z^{N_{\ua}+1}} E_3^N = \frac{1}{2^{N_{\ua}}} C_N^{N_{\ua}} \label{eq:p1}.
\ee
The probability distribution reduces to a binomial distribution function with only one peak located at $N_{\ua}^{(max)} = N/2$. Therefore, this distribution function corresponds to the unpolarized phase. 

For the case with $\theta \gg \omega$, $f_z \rar 0$, thus  the second term  proportional to $(1-f_z)$  in Eq.~(\ref{eq:Z0}) dominates. Near $z\sim 0$,  we can approximate  $\varepsilon(z)$ up to first order of $z$:  $\varepsilon = q_2 + q_1 z$, with $q_1 = \frac{\theta^2 + \omega^2}{\sqrt{\theta^2-\omega^2} }$ and $q_2 = \sqrt{\theta^2-\omega^2} $.  Then we obtain: 
\be
P(N_{\ua}, N)  = \frac{1}{2^{N+1}} C_N^{N_{\ua}} && \Big[  (1+q_1)^{N_{\ua}}  (1- q_2)^{N-N_{\ua}} \nn \\
&& +   (1-q_1)^{N_{\ua}}  (1+ q_2)^{N-N_{\ua}} \Big] \label{eq:p2}.
\ee
In this case, the distribution function is a combination of two binomial distribution function. In the large $N$ limit, it corresponds to two peaks located at
\be
N_{\ua}^{(max)} = \frac{1+q_1}{2+q_1 -q_2}N, ~~{\rm and }~~\frac{1-q_1}{2-q_1 +q_2}N,
\ee 
which are no longer $N/2$. 

The above two results are plotted in Fig.~\ref{distribution} together with that obtained from exact numerical calculations. It is seen that the analytical result agrees well with numerical results.

\section{Discussions}

Until now,  our studies are mainly focused on the situation with initial state  $\hrho_0 = \frac{1}{2}(\hat{1} + \hsigma_x)$ and with relaxation rate set to be zero.  To complete our studies, we would like to briefly discuss the effects of different initial quantum states and additional relaxation rate on the phase diagram. 

\begin{figure}[!htb]
 \scalebox{0.45}[0.45]{\includegraphics{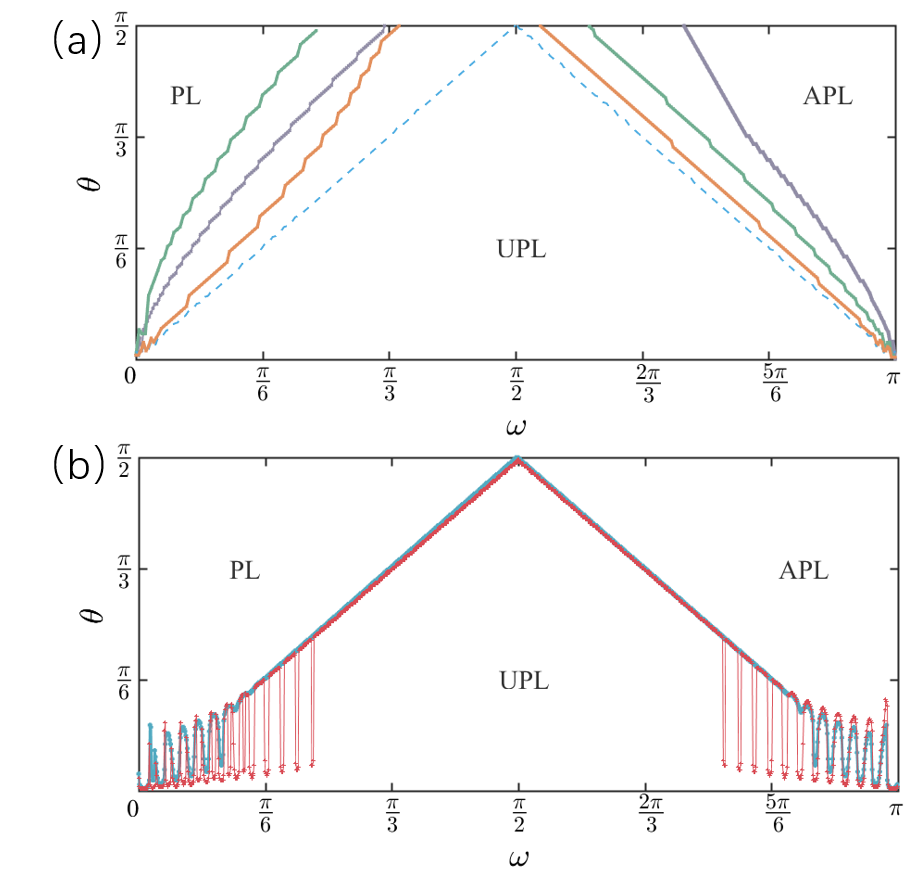}}
\caption{The phase diagram with different initial states (a) and different relaxation rates (b).  The initial states are, respectively, $\bp_0=(1, 1/2, \sqrt{3}/2)$ (red), $\bp_0=(1, \sqrt{3}/2, 1/2)$ (gray), and $\bp_0 = (1, 1, 0)$ (blue).  Here, the number of measurements is $N=1000$.  }
\label{fig4}
\end{figure}

First, we plot the phase diagram for different initial states in Fig.~\ref{fig4}(a). We see that the phase boundaries are strongly modified at nonzero $\omega$ for different initial states. This may be attributed to the fact that our criteria to determine  the boundary between polarized phase and unpolarized phase is too sensitive to the initial state. In contrast, the boundary between the unpolarized phase and anti-polarized phase is a little bit more robust, as shown by the red line.  Nevertheless,  deep inside the three phases, the probability distributions of  ferromagnetic order and anti-ferromagnetic order are still very well distinguished, indicating that the description of single qubit dynamics in terms of the language of  phase transition is still very useful. 

Secondly, we discuss the case with nonzero relaxation rate. The relaxation rate $r$ can be introduced initially in the quantum master equation: 
\be
\frac{d}{dt}{\hrho} = -\frac{i}{\hbar} [\hat{H}, \hrho] - r (\hrho - \hrho^{(th)} )
\ee
with $\hrho^{(th)} = \hat{1}/2$ being the density matrix for the totally thermal state. Repeating the same procedure as in Section II, we obtain that the evolving equation for state vector $\bp_n$ should be modified as: 
\be
{\cal A}_{\alpha} = \frac{1}{2}\left(
\begin{array}{ccc}
 1 & \alpha  \sin  \theta  \cos \omega e^{-r \tau} & -\alpha  \sin  \theta  \sin \omega  e^{-r \tau}  \\
 \alpha  \sin \theta  & \cos \omega  e^{-r \tau} & -\sin\omega  e^{-r \tau} \\
 0 & \cos  \theta  \sin\omega  e^{-r \tau}  & \cos  \theta  \cos \omega  e^{-r \tau}  \\
\end{array}
\right)  \nn \\
 \label{eq:Ar} 
\ee
Using the same recursion relations and Eq.~(\ref{eq:Ar}), we plot the phase diagram for the case of nonzero relaxation rate in Fig.~\ref{fig4}(b). We see that, if the relaxation rate is increased, the oscillation behavior becomes more amplified. However, as long as the relaxation rate is sufficiently small, it doesn't change the phase boundary.

\section{Conclusion}

In conclusion, in this paper, we have studied the phase transitions induced by quantum measurement on a single qubit that is precessing around an external magnetic field. The corresponding phase diagram is obtained numerically by a fast algorithm we developed. By resorting to a long-range interacting Ising model, and the full counting statistical approach, the phase diagram can be quantitatively understood. The presence of magnetic field serves as an additional degree of freedom, and can be easily achieved and controlled in the experiment. Our findings deepen the understanding of phase transition induced by quantum measurement, and may shed light on the characterization and monitoring of quantum state evolution \cite{Pfender2019, bcs} and find its future application in quantum tomography and quantum sensing. 

\section*{Acknowledgements}
F. Li was supported by NSFC (No. 11905054) and by the Fundamental Research Funds for the Central Universities from China. 


\end{document}